\documentclass[preprint,aps,showpacs,nofootinbib,tightenlines]{revtex4-1}
\usepackage{amsmath}
\usepackage{amssymb}
\usepackage{epsfig}
\usepackage{graphicx}
\begin{document}
%%\begin{CJK*}{GBK}{}
\newcommand{\beq}{\begin{eqnarray}}
\newcommand{\eeq}{\end{eqnarray}}

\newcommand{\non}{\nonumber\\ }
\newcommand{\ov}{\overline}
\newcommand{\rmt}{ {\rm T}}
\newcommand{\psl}{ p \hspace{-2.0truemm}/ }
\newcommand{\qsl}{ q \hspace{-2.0truemm}/ }
\newcommand{\epsl}{ \epsilon \hspace{-2.0truemm}/ }
\newcommand{\nsl}{ n \hspace{-2.2truemm}/ }
\newcommand{\vsl}{ v \hspace{-2.2truemm}/ }
\def \jpsi{ J/\Psi }

%%%%%%%%%%%%%%%%%%%
\def \cpc{{\bf Chin. Phys. C }}
\def \ctp{{\bf Commun.Theor.Phys. }}
\def \epjc{{\bf Eur.Phys.J. C}}
\def \jhep{ {\bf JHEP  }}
\def \jpg{ {\bf J.Phys. G}}
\def \npb{ {\bf Nucl.Phys. B}}
\def \plb{ {\bf Phys.Lett. B}}
\def \prd{ {\bf Phys.Rev. D}}
\def \prl{ {\bf Phys.Rev.Lett.}}
\def \ptp{ {\bf Prog. Theor. Phys.}}
\def \zpc{ {\bf Z.Phys.C}}

%%%%%%%%%%%%%%%%%%%%%%%%%%%%%%%%%%%%%%%%%%%%%%%%%%%%%%%%%%%

%%%%%%%%%%%%%%%%%%%%%%%%%%%%%%%%%%%%%%%%%%%%%%%%
\title{Semileptonic decays $B_c\to(\eta_c,J/\Psi)l\nu$ in the perturbative QCD approach}
\author{Wen-Fei Wang, Ying-Ying Fan and  Zhen-Jun Xiao\footnote{xiaozhenjun@njnu.edu.cn}}
\affiliation{
Department of Physics and Institute of Theoretical Physics,\\
Nanjing Normal University, Nanjing, Jiangsu 210023, People's Republic of China}
\date{\today}
%---------------------------------------------------------------------------------------------------------------%
\begin{abstract}
In this paper we study the semileptonic decays of $B_c^- \to (\eta_c,J/\psi)l^-\bar\nu_l$.
We firstly evaluate the $B_c \to (\eta_c,J/\Psi)$ transition
form factors $F_0(q^2)$, $F_+(q^2)$, $V(q^2)$ and $A_{0,1,2}(q^2)$ by employing the pQCD
factorization approach, and then we calculate the branching ratios for all considered
semileptonic decays. Based on the numerical results and the
phenomenological analysis, we find that:
(a) the pQCD predictions for the values of the $B_c \to \eta_c$ and $B_c \to J/\Psi$
transition form factors agree well with those obtained by using other methods;
(b) the pQCD predictions for the branching ratios of the considered decays are
$Br\left(B_c^- \to\eta_c e^-\bar\nu_e(\mu^-\bar\nu_\mu)\right)
=(4.41^{+1.22}_{-1.09})\times10^{-3}$,
$Br\left(B_c^- \to\eta_c\tau^-\bar\nu_\tau\right)
=(1.37^{+0.37}_{-0.34})\times10^{-3}$,
$Br(B_c^- \to J/\Psi e^-\bar\nu_e(\mu^-\bar\nu_\mu))
=(10.03^{+1.33}_{-1.18})\times10^{-3}$, and
$Br\left(B_c^- \to J/\Psi\tau^-\bar\nu_\tau\right)
=(2.92^{+0.40}_{-0.34})\times10^{-3}$;
and (c) we also define and calculate two ratios of the branching ratios
$R_{\eta_c}$ and $R_{J/\Psi}$, which will be tested by LHCb and the forthcoming
Super-B experiments.
\end{abstract}


\maketitle
%---------------------------------------------------------------------------------------------------------------%

\section{Introduction}\label{sec:1} %cccc

The charmed $B_c$ meson was found by the CDF Collaboration at Tevatron \cite{58-112004}
fifteen years ago.
It is a pseudoscalar ground state of two heavy quarks $b$ and $c$, which can decay individually.
Being below the $B-D$ threshold, $B_c$ meson  can only decay through weak interactions, it is
an ideal system to study weak decays of heavy quarks.
Due to the different decay rate of the two heavy quarks, the $B_c$ meson decays are rather different from those
of $B$ or $B_s$ meson.
Although the phase space in $c\to s$ transition is smaller
than that in $b\to c$ decays, but the Cabibbo-Kobayashi-Maskawa
(CKM) matrix element $|V_{cs}|\sim 1$ is much larger than $|V_{cb}|\sim 0.04$.
Thus the $c$-quark decays provide the dominant contribution (about $70\%$) to the decay width of
$B_c$ meson \cite{67-1559}.
At LHC experiments, around $5\times10^{10}$ $B_c$ events per year are expected
\cite{pap-LHC,67-1559}, which provide a very good platform to study various $B_c$ meson decay modes.

In fact, the $B_c$ decays have been studied intensively by many authors
\cite{79-034004,74-074008,71-094006,68-094020,77-114004,79-054012,1208-5916,81-014022,45-711,60-107}.
In Ref.\cite{79-034004}, for example, Dhir and Verma presented a detailed analysis
of the $B_c$ decays in the Bauer-Stech-Wirbel (BSW) framework,  while the authors of the Refs.~\cite{74-074008,71-094006,68-094020}
studied the $B_c$ meson decays in the non-relativistic or relativistic quark model.
In the perturbative QCD (pQCD) approach, furthermore, various $B_c$ decay modes have also
been studied for example in Refs.\cite{81-014022,45-711,60-107}.

\begin{figure}[tbp]
\vspace{-1cm}
\centerline{\epsfxsize=10cm \epsffile{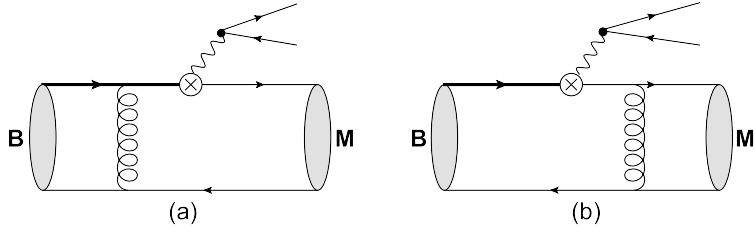}}
\caption{ The typical Feynman diagrams for the semileptonic
decays $B_c \to (\eta_c, J/\Psi)l\bar{\nu}$, where $B$ stands
for $B_c$ meson and $M$ for $\eta_c$ or $J/\Psi$ meson.}
\label{fig:fig1}
\end{figure}

In this paper, we will study the semileptonic decays of $B_c\to(\eta_c,J/\Psi)l\bar\nu_l$
(here $l$ stands for leptons $e, \mu,$ and $\tau$.) in the pQCD factorization
approach \cite{pap-pQCD,lu2001}.
The lowest order diagrams for $B_c\to(\eta_c,J/\Psi)l\bar\nu_l$ are displayed
in Fig.\ref{fig:fig1}, where $B$ stands
for $B_c$ meson and $M$ for $\eta_c$ or $J/\Psi$ meson, and the leptonic pairs
come from the $b$-quark's weak decay. In this work, we firstly calculate the $q^2$-dependent
form factors for $B_c\to (\eta_c,J/\Psi)$ transitions, and then we will give the branching ratios of the considered
semileptonic decay modes.

The structure of this paper is as below: after this introduction, we collect the distribution amplitudes of the
$B_c$ meson and the $\eta_c, J/\Psi$ mesons in Sec.II.
In Sec.III, based on the $k_{\rm T}$ factorization theorem, we calculate and present the expressions for the
$B_c \to (\eta_c, J/\Psi)$ transition form factors in the large recoil regions.
The numerical results and relevant discussions are given in Sec.~IV, a short summary will also be included in this
section.

%---------------------------------------------------------------------------------------------------------------%
\section{Kinematics and the wave functions}\label{sec:2}

In the $B_c$ meson rest frame, with the $m_{B_c}(m)$ stands
for the mass of the $B_c(\eta_c$ or $J/\Psi)$ meson, and $r=m/m_{B_c}$, the momenta of $B_c$ and $\eta_c(J/\Psi)$
mesons are defined in the same way as in Ref.~\cite{67-054028}
\beq
\label{eq-mom-p1p2}
p_1=\frac{m_{B_c}}{\sqrt{2}}(1,1,0_\bot),\quad
p_2=\frac{m_{B_c}}{\sqrt{2}}(r\eta^+,r\eta^-,0_\bot),
\eeq
where $\eta^+=\eta+\sqrt{\eta^2-1}$ and $\eta^-=\eta-\sqrt{\eta^2-1}$ with the definition of
the $\eta$ as of the form
\beq
\eta=\frac{m_{B_c}}{2m}\left[1+\frac{m^2-q^2}{m^2_{B_c}}\right],
\eeq
where $q^2= (p_1-p_2)^2$ is the invariant mass of the lepton pairs.
The momenta of the spectator quarks in $B_c$ and $\eta_c(J/\Psi)$ mesons are parameterized as
\beq\label{eq-k1k2}
k_1 =(0,x_1\frac{m_{B_c}}{\sqrt{2}},k_{1\bot}), \quad
k_2=(x_2\frac{m_{B_c}}{\sqrt{2}}r\eta^+,
x_2\frac{m_{B_c}}{\sqrt{2}}r\eta^-,k_{2\bot}).
\eeq
For the $J/\Psi$ meson, we define its polarization vector
$\epsilon$ as
\beq\label{eq-def-epsilon}
\epsilon_L=\frac{1}{\sqrt2}(\eta^+,-\eta^-,0_\bot), \quad
\epsilon_T=(0,0,1),
\eeq
where $\epsilon_L$ and $\epsilon_T$ denotes the longitudinal and transverse  polarization of the $J/\Psi$ meson.

%%%%%%%%%%%%%%%%%%%%%%%%%%%%%%%%%%%%%%%%%%%%%%%%%%
%%%%%%%%%%%%%%%%%%%%%%%%%%%%%%%%%%%%%%%%%%%%%%%%%%

In the calculations, one can ignore the $k_T$ contributions of $B_c$ meson.
Furthermore, one can assume that the $b$ and $c$ quark in $B_c$ meson are on the
mass shell approximately and treat its wave function as a $\delta$ function.
In this work, we use the same distribution amplitude for $B_c$ meson
as those used in Refs.\cite{81-014022,45-711}
\beq\label{eq-Bc-wavefun}
\Phi_{B_c}(x)=\frac{i}{\sqrt{2N_c}}[(\psl+m_{B_c})\gamma_{5}
\phi_{B_c}(x)]_{\alpha\beta}
\eeq
with
\beq\label{eq-Bc-wave}
\phi_{B_c}(x)= \frac{f_{B_c}}{2\sqrt{2N_c}}\delta(x-m_c/m_{B_c})
\eeq
where $m_c$ is the mass of $c$-quark.

For pseudoscalar meson $\eta_c$, the wave function is the form
of
\beq\label{eq-wf-etac}
\Phi_{\eta_c}(x)=\frac{i}{\sqrt{2N_c}}\gamma_5
[\psl\phi^v(x)+m_{\eta_c}\phi^s(x)].
\eeq
The twist-2 and twist-3 asymptotic distribution amplitudes,
$\phi^v(x)$ and $\phi^s(x)$,
can be written as \cite{73-094006,612-215} %%vvv
\beq
\phi^v(x)&=&9.58\frac{f_{\eta_{c}}}{2\sqrt{2N_c}}x(1-x)
\left[\frac{x(1-x)}{1-2.8x(1-x)}\right]^{0.7}\;, \non
\phi^s(x)&=&1.97\frac{f_{\eta_{c}}}{2\sqrt{2N_c}}
\left[\frac{x(1-x)}{1-2.8x(1-x)}\right]^{0.7}\;.
\eeq

As for the vector $J/\Psi$ meson, we take the wave function
as follows,
\beq
\Phi_{J/\Psi}^L(x) &=&\frac{1}{\sqrt{2N_c}}
\bigg\{m_{J/\Psi}\epsl_L \phi^L(x)+\epsl_L
\psl\phi^{t}(x) \bigg\}\;;\non
\Phi_{J/\Psi}^T(x) &=&\frac{1}{\sqrt{2N_c}}
\bigg\{m_{J/\Psi}\epsl_T \phi^V(x)+\epsl_T
\psl\phi^{T}(x) \bigg\}\;.
\eeq
And the  asymptotic distribution amplitudes of $J/\Psi$ meson read as
\cite{73-094006}
\beq
\phi^L(x)&=&\phi^T(x)=9.58\frac{f_{J/\Psi}}{2\sqrt{2N_c}}x(1-x)
\left[\frac{x(1-x)}{1-2.8x(1-x)}\right]^{0.7}\;,\non
\phi^t(x)&=&10.94\frac{f_{J/\Psi}}{2\sqrt{2N_c}}(1-2x)^2
\left[\frac{x(1-x)}{1-2.8x(1-x)}\right]^{0.7}\;,\non
\phi^V(x)&=&1.67\frac{f_{J/\Psi}}{2\sqrt{2N_c}}[1+(2x-1)^2]
\left[\frac{x(1-x)}{1-2.8x(1-x)}\right]^{0.7}\;.
\eeq
Here, $\phi^L$ and $\phi^T$ denote for the twist-2
distribution amplitudes, and $\phi^t$ and $\phi^V$ for the twist-3 distribution amplitudes.

%---------------------------------------------------------------------------------------------------------------%
\section{Form factors and  semileptonic decays}\label{sec:3}

As is well-known, the form factors of $B_{(s)}\to P,V,S$\footnote{Here $P,V,S$ denote the
pseudoscalar, vector and scalar meson respectively.} transitions have been
calculated by many authors in the pQCD factorization approach and other methods\cite{65-014007,pQCD-ff},
and the pQCD predictions are generally consistent with those from other methods.

The $B_c\to \eta_c$ form factors induced by vector currents are
defined in Refs.~\cite{67-054028,zpc29-637,npb592-3}
\beq
\langle \eta_c(p_2)|\bar{c}(0)\gamma_{\mu}b(0)|B_c(p_1)\rangle=
\left[(p_1+p_2)_{\mu}-\frac{m_{B_c}^2-m^2}{q^2}q_{\mu}\right]
F_+(q^2) +\frac{m_{B_c}^2-m^2}{q^2}q_{\mu}F_0(q^2),
\eeq
where $q=p_1-p_2$ is the momentum transfer to the lepton pairs,
and $m$ is the mass of $\eta_c$ meson.
In order to cancel the poles at $q^2=0$, $F_+(0)$ should be equal
to $F_0(0)$. For the sake of the calculation, it is convenient to
define the auxiliary form factors $f_1(q^2)$ and $f_2(q^2)$
\beq
\langle \eta_c(p_2)|\bar{c}(0)\gamma_{\mu}b(0)|B_c(p_1)\rangle=
f_1(q^2)p_{1\mu}+f_2(q^2)p_{2\mu}.
\eeq
In terms of $f_1(q^2)$ and $f_2(q^2)$ the form factor $F_+(q^2)$ and $F_0(q^2)$ read
\beq\label{eq-f+f0}
F_+(q^2)&=&\frac12\left[f_1(q^2)+f_2(q^2)\right], \notag\\
F_0(q^2)&=&\frac12 f_1(q^2)\left[1+\frac{q^2}{m_{B_c}^2-m^2}\right]
+\frac12 f_2(q^2)\left[1-\frac{q^2}{m_{B_c}^2-m^2}\right].
\eeq
The form factors $F_+$ and $F_0$ (or $f_1$ and $f_2$)
of the $B_c\to\eta_c$ transition are dominated by a single gluon exchange in the leading-order and in the large recoil regions.
And in the hard-scattering kernel, the transverse momentum
is retained to regulate the endpoint singularity.
The factorization formula for
the $B\to \eta_c$ form factors in pQCD approach
is written as \cite{65-014007}
\beq
\langle \eta_c(p_2)|\; \bar{c}(0)\gamma_\mu b(0)|B_c(p_1)\rangle &=& g_s^2 C_F N_c \int dx_1 dx_2 d^2k_{1{\rm T}}d^2k_{2{\rm T}}
\frac{dz^+d^2 z_{\rm T}}{(2\pi)^3}\frac{dy^-d^2 y_{\rm T}}{(2\pi)^3 } \non
&& \hspace{-2cm}\times e^{-ik_2\cdot y}\langle \eta_c(p_2)|\bar c_\gamma^\prime(y)
c_\beta(0)|0\rangle e^{i k_1\cdot z}\langle 0|\bar{c}_\alpha(0)
b_\delta^\prime(z)|B_c(p_1)\rangle T_{H\mu}^{\gamma\beta;\alpha\delta}.
\eeq

In the transverse configuration b-space and by
including the Sudakov form factors and the threshold resummation
effects, we obtain the $B\to \eta_c$ form factors $f_1(q^2)$
and $f_2(q^2)$ as following,
\beq  %----------------  f1 f2 ---------------%
f_1(q^2)&=&8\pi m_{B_c}^2rC_F\int dx_1 dx_2\int b_1 db_1 b_2 db_2
\;\phi_{B_c}(x_1)\non
&\times &\Bigl\{2\left[ \phi^s(x_2)-rx_2\phi^v(x_2)
 \right]\cdot h_1(x_1,x_2,b_1,b_2)\cdot \alpha_s(t_1)
 \exp\left [-S_{ab}(t_1) \right ] \non
&+& \left[4r_c\phi^s(x_2)-2r\phi^v(x_2)+
\frac{x_1\eta^+(\eta^+\phi^v(x_2)-2\phi^s(x_2))}
{\sqrt{\eta^2-1}}
\right] \non
&\times& h_2(x_1,x_2,b_1,b_2)
\cdot \alpha_s (t_2)\exp\left [-S_{ab}(t_2) \right] \Bigr \},
\label{eq:f1q2}
\eeq
\beq
f_2(q^2)&=&8\pi m_{B_c}^2C_F\int dx_1 dx_2\int b_1 db_1 b_2 db_2 \;\phi_{B_c}(x_1)\non
&\times& \Bigl\{ \left[2\phi^v(x_2)-
4rx_2(\phi^s(x_2)-\eta\phi^v(x_2))\right]\non
&\times & h_1(x_1,x_2,b_1,b_2)\cdot \alpha_s(t_1)
 \exp\left [-S_{ab}(t_1) \right ] \non
&+& \left[ 4r\phi^s(x_2)-2r_c\phi^v(x_2)-
\frac{x_1(\eta^+\phi^v(x_2)-2\phi^s(x_2))}
{\sqrt{\eta^2-1}}
\right] \non
&\times & h_2(x_1,x_2,b_1,b_2)
\cdot \alpha_s (t_2)\exp\left [-S_{ab}(t_2) \right] \Bigr \},
\label{eq:f2q2}
\eeq %----------------  f1 f2 ------END------%
where $C_F=4/3$ is a color factor, $r$ is the same as in
Eqs.(\ref{eq-mom-p1p2},\ref{eq-k1k2}), while $r_c=m_c/m_{B_c}$, and $m_c$ is the mass of $c$-quark.
The functions $h_1$ and $h_2$, the scales $t_1$, $t_2$ and the Sudakov factors $S_{ab}$ are given in
the Appendix A of this paper.

For the charged current $B_c\to \eta_c l\bar\nu_l$, the quark level transitions are the
$b\to cl\bar\nu_l$ transition with the effective Hamiltonian
\cite{rmp68-1125}
\beq  \label{eq-hamiltonian}
{\cal H}_{eff}(b\to cl\bar \nu_l)=\frac{G_F}{\sqrt{2}}V_{cb}\;
\bar{c} \gamma_{\mu}(1-\gamma_5)b \cdot \bar l\gamma^{\mu}(1-\gamma_5)\nu_l,
\eeq
where $G_F=1.166 37\times10^{-5} GeV^{-2}$ is the Fermi-coupling constant.
With the two form factors $F_+(q^2)$ and $F_0(q^2)$, we can write down the differential decay width of the decay mode
$B_c\to \eta_c l\bar\nu_l$ as
\cite{73-115006,79-054012}
\beq
\frac{d\Gamma(B_c\to \eta_cl\bar \nu_l)}{dq^2}&=&\frac{G_F^2|V_{cb}|^2}{192 \pi^3
 m_{B_c}^3}\frac{q^2-m_l^2}{(q^2)^2}\sqrt{\frac{\left (q^2-m_l^2 \right )^2}{q^2}}
 \sqrt{\frac{\left (m_{B_c}^2-m^2-q^2 \right )^2}{4q^2}-m^2}\non
&& \times \Bigl \{ \left (m_l^2+2q^2 \right ) \left [q^2-\left (m_{B_c}-m \right )^2 \right ]
\left [ q^2-\left (m_{B_c}+m \right )^2 \right ]F_+^2(q^2) \non
&& + 3m_l^2\left (m_{B_c}^2-m^2 \right )^2 F_0^2(q^2)\Bigr \},
 \eeq
where $m_l$ and $m$ is the mass of the lepton and $\eta_c$ respectively.
If the produced lepton is $e^{\pm}$ or $\mu^{\pm}$, the corresponding mass terms of the lepton could be neglected.

The form factors $V(q^2)$ and $A_{0,1,2}(q^2)$ for $B_c\to J/\Psi$ transition are defined in the same way as
in Refs.~\cite{67-054028,zpc29-637,npb592-3} and  are written explicitly as,
\beq  %----------------  V A_{1,2,3} ---------------%
V(q^2)&=&4\pi m_{B_c}^2C_F\int dx_1 dx_2\int b_1 db_1 b_2 db_2
 \;\phi_{B_c}(x_1)\cdot (1+r)\non
&\times & \Bigl \{2\left[\phi^T(x_2)-rx_2\phi^V(x_2)\right]
\cdot h_1(x_1,x_2,b_1,b_2)\cdot \alpha_s(t_1)
 \exp\left [-S_{ab}(t_1) \right ] \non
&+& \left[\left(2r+\frac{x_1}{\sqrt{\eta^2-1}}\right)\phi^V(x_2)
\right]\cdot h_2(x_1,x_2,b_1,b_2)
\cdot \alpha_s (t_2)\exp\left [-S_{ab}(t_2) \right] \Bigr \},
\label{eq:Vqq}
\eeq
\beq
A_0(q^2)&=&8\pi m_{B_c}^2C_F\int dx_1 dx_2\int b_1 db_1 b_2 db_2
 \;\phi_{B_c}(x_1)\non
&\times & \Bigl \{  \left[\left(1-r^2x_2+2rx_2\eta\right)\phi^L(x_2)
+r\left(1-2x_2\right)\phi^t(x_2) \right]\non
&\times& h_1(x_1,x_2,b_1,b_2)\cdot \alpha_s(t_1)
 \exp\left [-S_{ab}(t_1) \right ] \non
&+& \left [\left(r^2+r_c+\frac{x_1}{2}-rx_1\eta
+\frac{x_1(\eta+r(1-2\eta^2))}{2\sqrt{\eta^2-1}}
 \right)\phi^L(x_2)\right ]\non
&\times & h_2(x_1,x_2,b_1,b_2)
\cdot \alpha_s (t_2)\exp\left [-S_{ab}(t_2) \right] \Bigr \},
\label{eq:A0qq}
\eeq
\beq
A_1(q^2)&=&8\pi m_{B_c}^2C_F\int dx_1 dx_2\int b_1 db_1 b_2 db_2
\;\phi_{B_c}(x_1)\cdot \frac{r}{1+r}\non
&\times & \Bigl \{\left [ 2(1+rx_2\eta)\phi^V(x_2)
-2(2rx_2-\eta)\phi^T(x_2) \right ]\non
&\times& h_1(x_1,x_2,b_1,b_2)\cdot \alpha_s(t_1)\exp[-S_{ab}(t_1)]\non
& + &\left[ \left(2r_c-x_1+2r\eta \right) \phi^V(x_2)\right]\cdot h_2(x_1,x_2,b_1,b_2)
\cdot \alpha_s (t_2)\exp[-S_{ab}(t_2)] \Bigr \},
\label{eq:A1qq}
\eeq
\beq
A_2(q^2)&=&\frac{(1+r)^2(\eta-r)}{2r(\eta^2-1)}\cdot A_1(q^2)-
8\pi m_{B_c}^2C_F\int dx_1 dx_2\int b_1 db_1 b_2 db_2
\;\phi_{B_c}(x_1)\cdot \frac{1+r}{\eta^2-1}\non
&\times & \Bigl \{  \left[(\eta(1-r^2x_2)-r(1+x_2-2x_2\eta^2))
\phi^L(x_2)
+\left(1+2r^2x_2-r\eta(1+2x_2)\right)\phi^t(x_2)\right]\non
&\times&
h_1(x_1,x_2,b_1,b_2)\cdot \alpha_s(t_1)
 \exp\left [-S_{ab}(t_1) \right ] \non
&+& \left[ x_1\left(r\eta-\frac12\right)\sqrt{\eta^2-1}
+\left(r_c+r^2-\frac{x_1}{2}\right)\eta
+r\left(1-r_c-\frac{x_1}{2}+x_1\eta^2\right)
\right]\phi^L(x_2)\non
&\times& h_2(x_1,x_2,b_1,b_2)
\cdot \alpha_s (t_2)\exp\left [-S_{ab}(t_2) \right] \Bigr \},
\label{eq:A2qq}
\eeq
wherte $r=m_{J/\Psi}/m_{B_c}$, $C_F$ and $r_c$ are the same as in Eqs.(\ref{eq:f1q2},\ref{eq:f2q2}).
The expressions of the hard function $h_1$ and $h_2$, hard scales $t_1$ and $t_2$, and Sudakov function $S_{ab}$
are all given in the Appendix A.
One should note that the pQCD predictions for the form factors $f_{1,2}(q^2)$, $V(q^2)$ and $A_{0,1,2}(q^2)$
as given in Eqs.(\ref{eq:f1q2},\ref{eq:f2q2},\ref{eq:Vqq}-\ref{eq:A2qq})
are all leading order results. The NLO contributions  to the form factors of $B \to (\pi, K)$ transitions as
given in Refs. \cite{85-074004,86-114025} do not apply here because of the large mass of $c$-quark and
$(\eta_c, J/\Psi)$ mesons.

The effective Hamiltonian for the decay modes
$B_c\to J/\Psi l\bar\nu_l$ is the same as $B_c\to \eta_c l\bar\nu_l$,
but corresponding differential decay widths are different.
For $B_c\to J/\Psi l\bar\nu_l$, we have\cite{79-054012,JHEP10206,83-032007}
\beq
\frac{d\Gamma_L(B_c\to J/\Psi l\bar \nu_l)}{dq^2}&=&\frac{G_F^2|V_{cb}|^2}{192 \pi^3
 m_{B_c}^3}\frac{q^2-m_l^2}{(q^2)^2}\sqrt{\frac{\left (q^2-m_l^2 \right )^2}{q^2}}
 \sqrt{\frac{\left (m_{B_c}^2-m^2-q^2 \right )^2}{4q^2}-m^2}\non
&\times & \Bigg\{
3m^2_l\lambda(q^2)A^2_0(q^2)+\frac{m^2_l+2q^2}{4m^2}\non
&\times &\Bigg[(m^2_{B_c}-m^2-q^2)(m_{B_c}+m)A_1(q^2)
 -\frac{\lambda(q^2)}{m_{B_c}+m}A_2(q^2) \Bigg]^2 \Bigg\},
\eeq
\beq
\frac{d\Gamma_\pm(B_c\to J/\Psi l\bar \nu_l)}{dq^2}&=&\frac{G_F^2|V_{cb}|^2}{192 \pi^3
 m_{B_c}^3}\frac{q^2-m_l^2}{q^2}\sqrt{\frac{\left (q^2-m_l^2 \right )^2}{q^2}}
 \sqrt{\frac{\left (m_{B_c}^2-m^2-q^2 \right )^2}{4q^2}-m^2}\non
& \times & \Bigg \{ (m^2_l+2q^2)\lambda(q^2)
\left[\frac{V(q^2)}{m_{B_c}+m}\mp
\frac{(m_{B_c}+m)A_1(q^2)}{\sqrt{\lambda(q^2)}}\right]^2\Bigg\},
\eeq
where $m=m_{\jpsi}$, and $\lambda(q^2) = (m_{B_c}^2+m^2-q^2)^2 - 4 m_{B_c}^2 m^2$ is the phase-space factor.
The combined transverse and total differential decay widths are defined as
\beq
\frac{d\Gamma_T}{dq^2}=\frac{d\Gamma_+}{dq^2}   +\frac{d\Gamma_-}{dq^2} \; ,\quad
\frac{d\Gamma}{dq^2}=\frac{d\Gamma_L}{dq^2}   +\frac{d\Gamma_T}{dq^2} \; .
\eeq

%---------------------------------------------------------------------------------------------------------------%
\section{Numerical results and discussions} \label{sec:4}

In the numerical calculations we use the following input parameters (here masses and decay
constants in unit GeV ) \cite{81-014022,pdg2012}
\beq
\Lambda^{(f=4)}_{\bar{MS}}&=&0.287, \quad m_{\eta_c}=2.981, \quad
m_{J/\Psi}=3.097, \quad m_{B_c}=6.277,\non
m_{c}&=&1.275\pm0.025, \quad m_{\tau}=1.777,\quad f_{B_c}=0.489, \quad
\tau_{B_c}=(0.45\pm0.04)\; ps,\non
|V_{cb}|&=&(41.2^{+1.1}_{-0.5})\times 10^{-3},\quad
f_{\eta_c}=(0.420\pm 0.050),\quad
f_{J/\Psi}=(0.405\pm 0.014).
\label{eq:inputs}
\eeq

%%%%%%%%%%%%%%%%%%%%%%%%%%%%%%%%%%%%%%%%%%%%%%%%
\begin{figure}
\begin{center}
\vspace{-1cm}
\centerline{\epsfxsize=5cm \epsffile{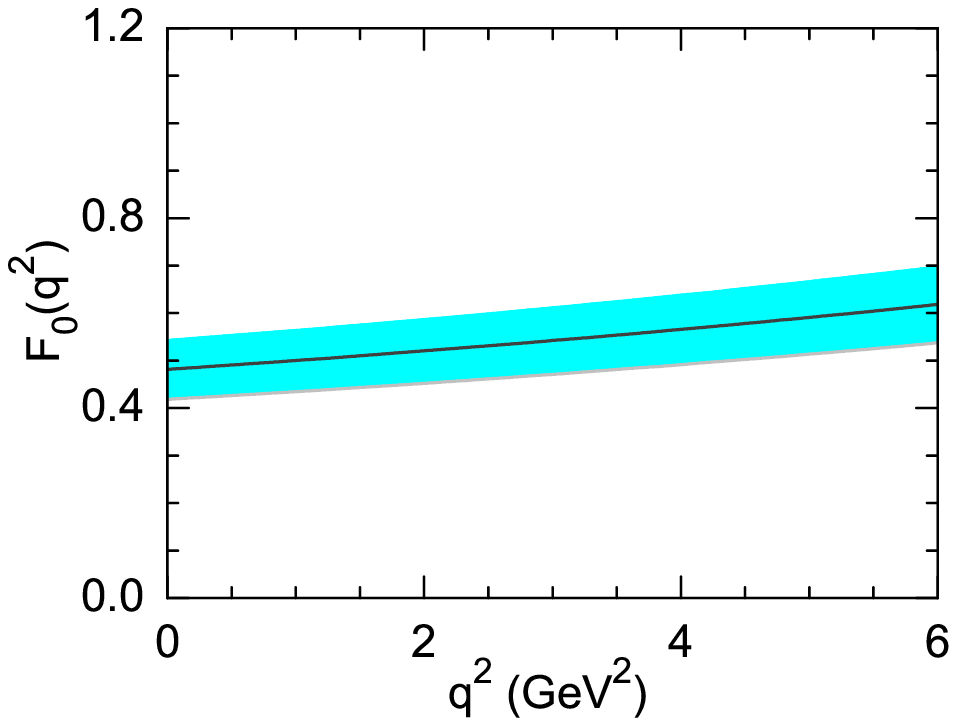}
\epsfxsize=5cm \epsffile{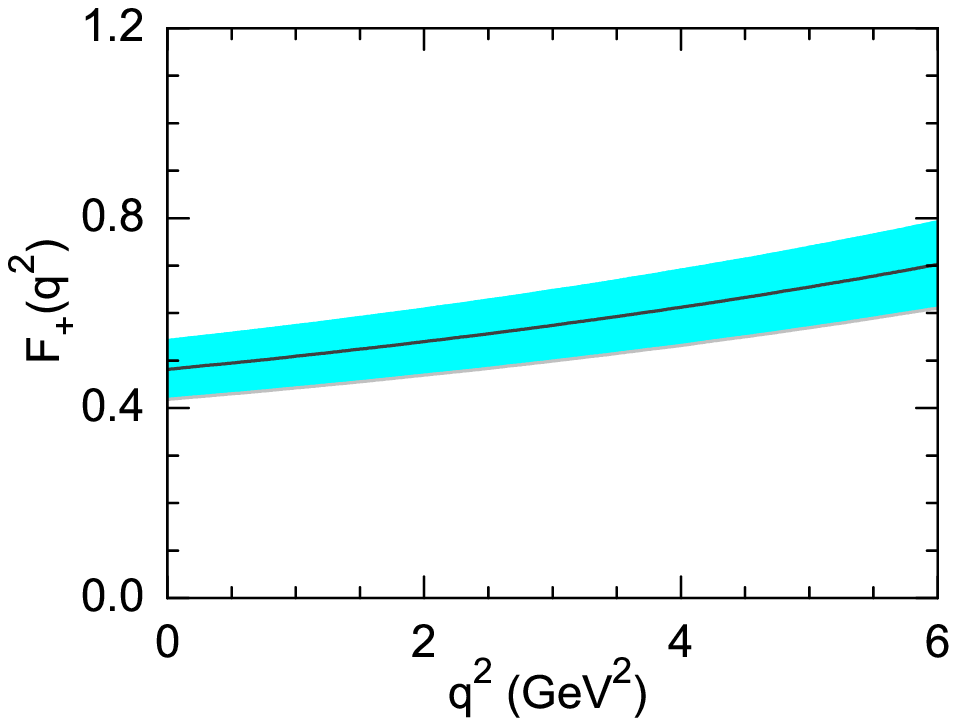}
\epsfxsize=5cm \epsffile{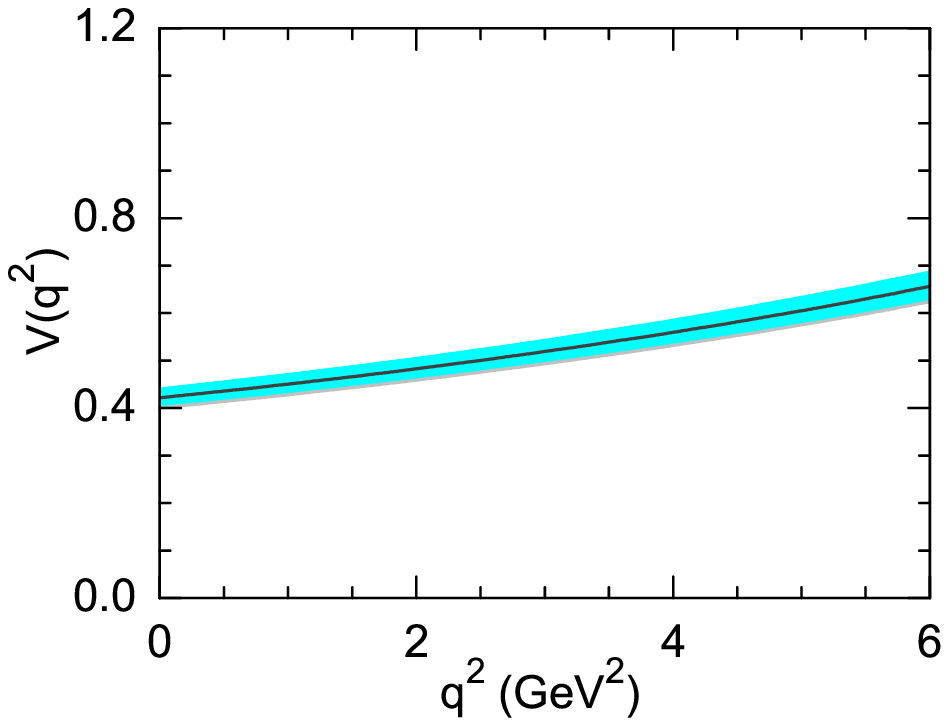}}
\centerline{\epsfxsize=5cm \epsffile{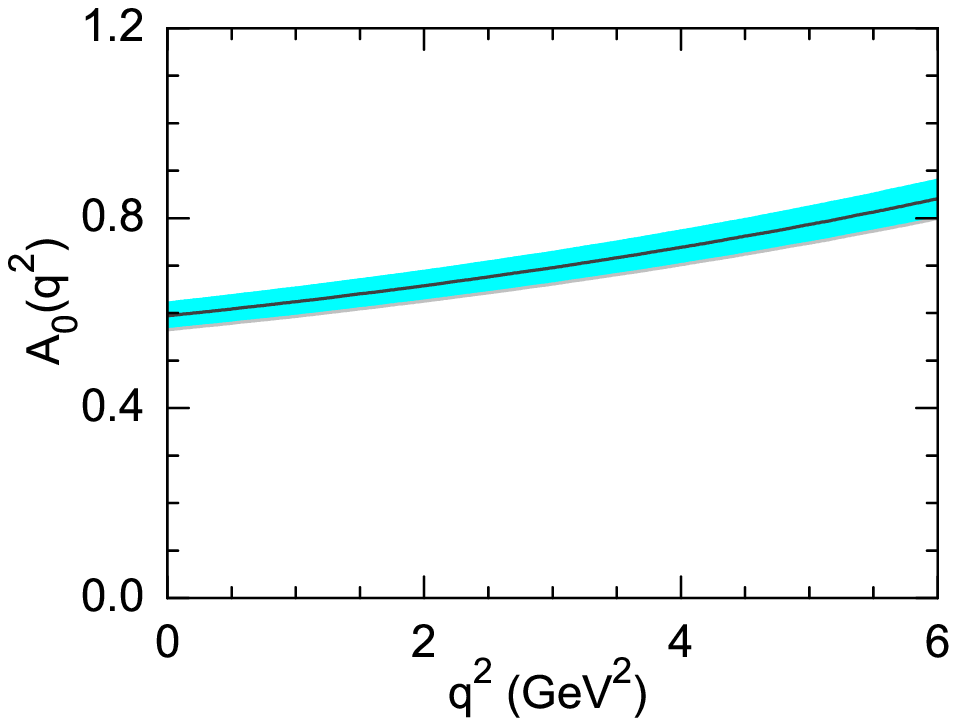}
\epsfxsize=5cm \epsffile{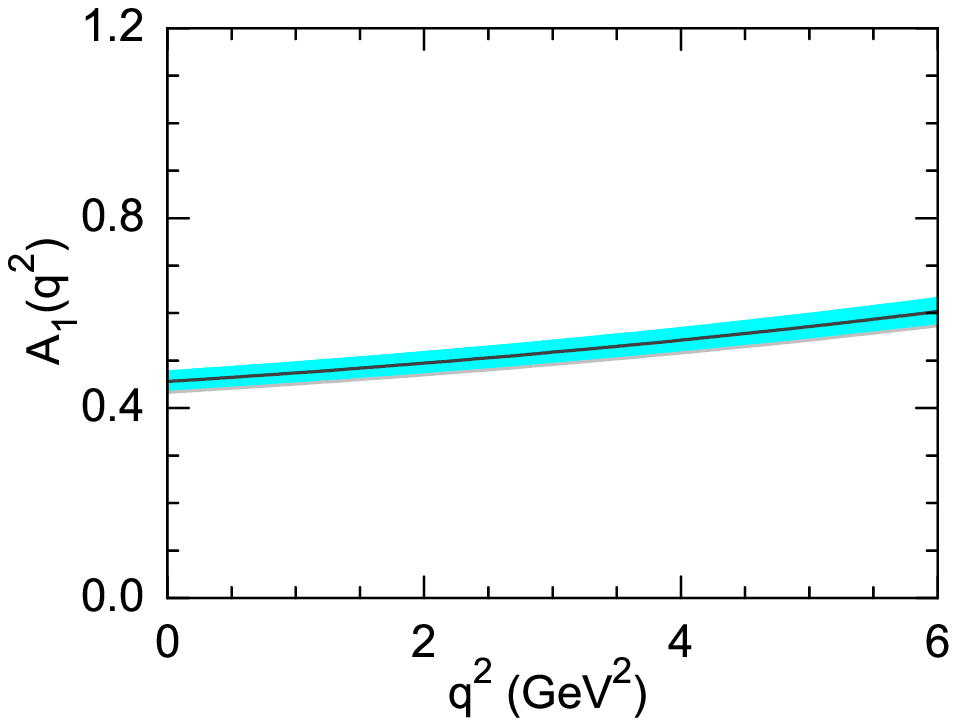}
\epsfxsize=5cm \epsffile{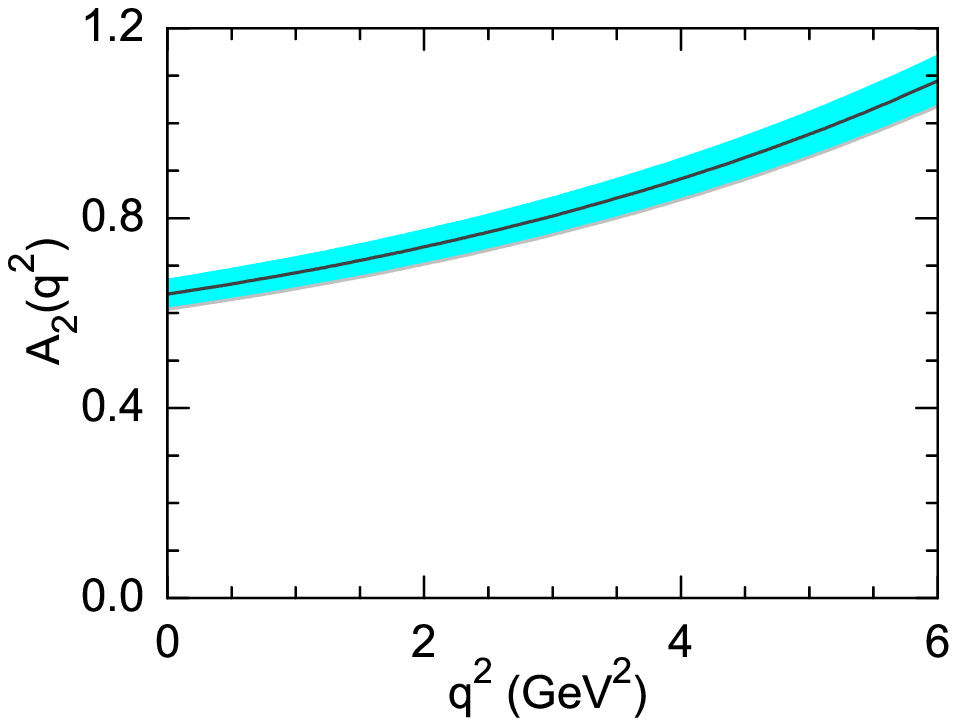}}
\caption{The pQCD predictions for the $q^2$-dependence
of the form factors $F_0, F_+, V, A_0, A_1$ and $A_2$.
With the solid lines stand for the central values, and the
bands show the errors of the corresponding form factors.}
\label{fig:fig2}
\end{center}
\end{figure}
%%%%%%%%%%%%%%%%%%%%%%%%%%%%%%%%%%%%%%%%%%%%%%%%%%

By using the expressions in Eqs.(\ref{eq:f1q2},\ref{eq:f2q2},\ref{eq:Vqq}-\ref{eq:A2qq})
and the definitions in Eq.~(\ref{eq-f+f0})
we calculate the values of the form factors $F_0(q^2)$, $F_+(q^2)$,
$V(q^2)$, $A_0(q^2)$, $A_1(q^2)$ and $A_2(q^2)$
for given value of $q^2$ in the region
of $0\leq q^2 \leq (m_{B_c}-m)^2$.
But one should note that the pQCD predictions for the
considered form factors are reliable only for small
values of $q^2$. For the form factors in the larger $q^2$ region,
one has to make an extrapolation for them from the lower $q^2$
region to larger $q^2$ region. In this work we make the
extrapolation by using the formula in Refs.~\cite{79-054012,1209.1157}
\beq
F(q^2)=F(0)\cdot \exp{\left[a\cdot q^2+b\cdot (q^2)^2\right]}.
\eeq
where $F$ stands for the form factors $F_{0,+}, V, A_{0,1,2}$,
and $a, b$ are the parameters to be determined by the fitting procedure.

The numerical values of the form factors $F_{0,+}, V$ and $A_{0,1,2}$  at $q^2=0$
and their fitted parameters are listed in Table \ref{tab-FF}.
The first error of the pQCD predictions for the form factors in Table \ref{tab-FF}
comes from the uncertainty of the decay constants
of $\eta_c$ and/or $J/\Psi$ mesons, and the second error comes
from the uncertainty of $m_c=1.275\pm0.025$ \cite{pdg2012}.
For the parameters $(a, b)$, their errors from the
decay constants $f_{\eta_c}, f_{J/\Psi}$ or from $m_c=1.275\pm0.025$
are negligibly small and not shown here explicitly.
As a comparison, we also present some results obtained by other authors
based on different methods
in Table \ref{tab-FF-literature}.
One should note that the definition of the $B_c\to J/\Psi$ transition form factors in this paper
are different from those in Ref.~\cite{71-094006} (ISK) and \cite{74-074008} (HNV).
From Table \ref{tab-FF-literature}, we find
that our results agree well with the results in other literatures.
In Fig.\ref{fig:fig2}, we show the pQCD predictions for
the $q^2$-dependence of those form factors, where the
solid lines stand for the central values, and the bands show the
theoretical  errors of the corresponding form factors.

%--------------------------------------------------
\begin{table}[thb]
\begin{center}
\caption{The pQCD predictions for form factors $F_{0,+}, V, A_{0,1,2}$ at $q^2=0$
and the parametrization constants $a$ and $b$
for $B_c\to \eta_c$ and $B_c \to J/\Psi$ transitions.}
\label{tab-FF}
\begin{tabular}{c| c c c} \hline\hline
\ \ \ &$ \ \ F(0)$ &$\ \ a$ &$ \ \ b$  \\
 \hline\hline
\ \ \ $F_0^{B_c\to\eta_c}$ &$0.48\pm0.06\pm0.01$&\ \ $0.037$
& \ \  $0.0007$ \; \\  \hline
\ \ \
$F_+^{B_c\to\eta_c}$   &$0.48\pm0.06\pm0.01$ &\ \ $0.055$
&\ \ $0.0014$ \; \\   \hline\hline
 \ \ \
$V^{B_c\to J/\Psi}$   &$0.42\pm0.01\pm0.01$ &\ \ $0.065$
 &\ \ $0.0015$ \; \\ \hline
\ \ \
$A_0^{B_c\to J/\Psi}$ &$0.59\pm0.02\pm0.01$
&\ \ $0.047$ &\ \ $0.0017$ \; \\  \hline
\ \ \
$A_1^{B_c\to J/\Psi}$   &$0.46\pm0.02\pm0.01$
&\ \ $0.038$ &\ \ $0.0015$ \; \\   \hline
 \ \ \
$A_2^{B_c\to J/\Psi}$   &$0.64\pm0.02\pm0.01$
&\ \ $0.064$  &\ \ $0.0041$ \;  \\  \hline\hline
\end{tabular}
\end{center}
\end{table}
%---------------------------------------------------
\begin{table}[thb]
\begin{center}
\caption{$B_c\to \eta_c, J/\Psi$ transition form factors
at $q^2=0$ evaluated in this paper and in other literatures.}
\label{tab-FF-literature}
\begin{tabular}{l| c| c| c| c| c| c} \hline\hline
\ \ \ &$\ \textrm{pQCD}$ &$\textrm{WSL\cite{79-054012}}$&$\textrm{EFG\cite{68-094020}}$
 &$\textrm{ISK\cite{71-094006}}$
 &$\textrm{HNV\cite{74-074008}}$
  &$\textrm{DV\cite{79-034004}}$
\\ \hline\hline
$F_0^{B_c\to\eta_c}=
F_+^{B_c\to\eta_c}$    &$0.48$ &$0.61$ &$0.47$ &$0.61$ &$0.49$ &$0.58$
\\   \hline\hline
 \ \ \
$V^{B_c\to J/\Psi}$  \; &$0.42$ &$0.74$ &$0.49$ &$0.83$ &$0.61$ &$0.91$
\\ \hline
\ \ \
$A_0^{B_c\to J/\Psi}$\; &$0.59$ &$0.53$ &$0.40$ &$0.57$ &$0.45$ &$0.58$
\\  \hline
\ \ \
$A_1^{B_c\to J/\Psi}$\; &$0.46$ &$0.50$ &$0.50$ &$0.56$ &$0.49$ &$0.63$
\\   \hline
 \ \ \
$A_2^{B_c\to J/\Psi}$\; &$0.64$ &$0.44$ &$0.73$ &$0.54$ &$0.56$ &$0.74$
\\  \hline\hline
\end{tabular}
\end{center}
\end{table}
%--------------------------------------------------
%%%%%%%%%%%%%%%%%%%%%%%%%%%%%%%%%%%%%%%%%%%%%%%%

By using the relevant formulas and the input parameters as defined or
given in previous sections, it is straightforward to calculate the
branching ratios for all the considered decays.
By making the numerical integration over the physical range of
$q^2$, we find the pQCD predictions for the branching ratios of considered decay modes:
\beq
Br\left(B_c^- \to\eta_c e^-\bar\nu_e(\mu^-\bar\nu_\mu)\right)
&=&\left(4.41^{+1.11}_{-0.99}(f_{\eta_c})\pm0.39(\tau_{B_c})
^{+0.24}_{-0.11}(V_{cb})^{+0.22}_{-0.21}(m_c) \right)\times 10^{-3},\non
Br\left(B_c^- \to\eta_c\tau^-\bar\nu_\tau\right)
&=&\left(1.37^{+0.34}_{-0.31}(f_{\eta_c})\pm0.12(\tau_{B_c})
^{+0.07}_{-0.03}(V_{cb})^{+0.07}_{-0.06}(m_c) \right)\times 10^{-3},\non
Br\left(B_c^- \to J/\Psi e^-\bar\nu_e(\mu^-\bar\nu_\mu)\right)
&=&\left(10.03^{+0.71}_{-0.68}(f_{J/\Psi})\pm0.89(\tau_{B_c})
^{+0.54}_{-0.24}(V_{cb})^{+0.41}_{-0.27}(m_c) \right)\times 10^{-3},\non
Br\left(B_c^- \to J/\Psi \tau^-\bar\nu_\tau\right)
&=&\left(2.92^{+0.21}_{-0.20}(f_{J/\Psi})\pm0.26(\tau_{B_c})
^{+0.16}_{-0.07}(V_{cb})^{+0.12}_{-0.08}(m_c) \right)\times 10^{-3}.
\label{eq:brs}
\eeq
where the major theoretical errors come from the uncertainties of the input parameters
$f_{\eta_c}$, $f_{J/\Psi}$, $|V_{cb}|$, $\tau_{B_c}$ and $m_c$
as given explicitly in Eq.~(\ref{eq:inputs}).

From the pQCD predictions for the form factors $F_{0,+}, V, A_{0,1,2}$ as given in Table I
and the pQCD predictions for the
branching ratios of $B_c\to(\eta_c,J/\Psi)l\nu$  as given in Eq.(\ref{eq:brs}),
we find the following points:
\begin{enumerate}
\item[(i)]
The form factor $F_0(0)$ equals to $F_+(0)$ by definition, but they have
different $q^2$-dependence. The error bands of $F_0(q^2)$
and $F_+(q^2)$ in Fig.\ref{fig:fig2} are larger than that
of $V(q^2)$ and $A_{0,1,2}(q^2)$. The reason is that
the uncertainty of the decay constant $f_{\eta_c}$ in $B_c\to\eta_c$ transition is larger
than the one of $f_{J/\Psi}$ in $B_c\to J/\Psi$ transition.

\item[(ii)]
The pQCD predictions for the form factors  as listed in
Table \ref{tab-FF-literature} agree well with those obtained by using other methods.

\item[(iii)]
The pQCD predictions for the branching ratios of the four decay modes
$B_c\to(\eta_c,J/\Psi)l\nu$ are
at the order of $10^{-3}$. Because of its large mass of $\tau$ lepton, the
decays involving a $\tau$ in the final state has a smaller decay
rate than those with light $e^-$ or $\mu^-$.
Since the ratio of the branching ratios has
smaller theoretical error than the decay rates themselves, we here define
two ratios $R_{\eta_c}$ and $R_{J/\Psi}$, the pQCD predictions for them are the following
\beq
R_{\eta_c}&=&\frac{Br\left(B_c^- \to\eta_c l^-\bar\nu_l\right)}
{Br\left(B_c^- \to\eta_c\tau^-\bar\nu_\tau\right)}\approx 3.2, \quad for \quad l=(e,\mu), \\
R_{J/\Psi}&=&\frac{Br\left(B_c^- \to J/\Psi l^-\bar\nu_l\right)}
{Br\left(B_c^- \to J/\Psi\tau^-\bar\nu_\tau\right)}\approx 3.4, \quad for \quad l=(e,\mu).
\eeq
These relations will be tested by LHCb and the forthcoming Super-B experiments.
\end{enumerate}

In short we calculated the branching ratios of the semileptonic
decays $B_c^- \to (\eta_c,J/\psi)l^-\bar\nu_l$
in the pQCD factorization approach.
We first calculated the relevant form factors by employing the
pQCD factorization approach, and then evaluated  the branching
ratios for all considered semileptonic $B_c$ decays.
Based on the numerical results and the phenomenological analysis, we find that
\begin{enumerate}
\item[(i)]
For $B_c \to (\eta_c, J/\Psi)$ transitions, the LO pQCD predictions for the
form factors $F_{0,+}(0), V(0)$ and $A_{0,1,2}(0)$ agree with
those derived by using other different methods.

\item[(ii)]
The pQCD predictions for the branching ratios
of the considered decay modes are:
\beq
Br\left(B_c^- \to\eta_c e^-\bar\nu_e(\mu^-\bar\nu_\mu)\right)
&=&(4.41^{+1.22}_{-1.09})\times10^{-3},\non
Br\left(B_c^- \to\eta_c\tau^-\bar\nu_\tau\right)
&=&(1.37^{+0.37}_{-0.34})\times10^{-3},\non
Br(B_c^- \to J/\Psi e^-\bar\nu_e(\mu^-\bar\nu_\mu))
&=&(10.03^{+1.33}_{-1.18})\times10^{-3}, \non
Br\left(B_c^- \to J/\Psi\tau^-\bar\nu_\tau\right)
&=&(2.92^{+0.40}_{-0.34})\times10^{-3},
\eeq
where the individual theoretical errors in Eq.(\ref{eq:brs}) have been added in quadrature.

\item[(iii)]
We also defined two ratios of the branching ratios
$R_{\eta_c}$ and $R_{J/\Psi}$ and presented the corresponding
pQCD predictions, which will be tested by LHCb and the
forthcoming Super-B experiments.
\end{enumerate}

%---------- aaa ---------------------------------------%

\begin{acknowledgments}
This work is supported in part by the National Natural Science Foundation of
China under Grant No. 10975074 and 11235005.
\end{acknowledgments}

%---------------------------------------------------------------%

\appendix

\section{Relevant functions}

In this appendix, we present the functions appeared in the previous sections.
The threshold resummation factors $S_t(x)$ is adopted from Ref.~\cite{65-014007}:
\beq\label{eq-def-stx}
S_t=\frac{2^{1+2c}\Gamma(3/2+c)}{\sqrt{\pi}\Gamma(1+c)}[x(1-x)]^c,
\eeq
and we here set the  parameter $c=0.3$.
The hard functions $h_1$ and $h_2$ come form the Fourier transform
and can be written as
\beq
\begin{aligned}
h_1(x_1,x_2,b_1,b_2)&=K_0(\beta_1 b_1)
[\theta(b_1-b_2)I_0(\alpha_1b_2)K_0(\alpha_1b_1)\\
&+\theta(b_2-b_1)I_0(\alpha_1b_1)K_0(\alpha_1b_2)]S_t(x_2),
\end{aligned}
\eeq
\beq
\begin{aligned}
h_2(x_1,x_2,b_1,b_2)&=K_0(\beta_2 b_2)
[\theta(b_1-b_2)I_0(\alpha_2b_2)K_0(\alpha_2b_1)\\
&+\theta(b_2-b_1)I_0(\alpha_2b_1)K_0(\alpha_2b_2)]S_t(x_1),
\end{aligned}
\eeq
with $\alpha_1= m_{B_c}\sqrt{x_2r\eta}$, $\beta_1= m_{B_c}\sqrt{x_1x_2r\eta^+}$,
$\alpha_2=m_{B_c}\sqrt{x_1r\eta^+}$ and $\beta_2= \beta_1$.
The functions $K_0$ and $I_0$ are modified Bessel functions.

The factor $exp[-S_{ab}(t)]$ contains the Sudakov logarithmic corrections and
the renormalization group evolution effects of both the wave functions and the
hard scattering amplitude with $S_{ab}(t)=S_{B_c}(t)+S_M(t)$, where
\beq
S_{B_c}(t)&=&s(x_1\frac{m_{B_c}}{\sqrt{2}},b_1)+2\int_{1/b_1}^{t}
\frac{d\bar{\mu}}{\bar{\mu}}\gamma_q(\alpha_s(\bar{\mu})),\\
S_M(t)&=&s(x_2\frac{m_{B_c}}{\sqrt{2}},b_2)+s((1-x_2)\frac{m_{B_c}}{\sqrt{2}},b_2)
+2\int_{1/b_2}^{t}\frac{d\bar{\mu}}{\bar{\mu}}\gamma_q(\alpha_s(\bar{\mu})),
\eeq
with the quark anomalous dimension $\gamma_q=-\alpha_s/\pi$.
The explicit expressions of the functions $s(Q,b)$ can be found for example in Appendix A of
Ref.~\cite{lu2001}. The hard scales $t_i$ in Eqs.(\ref{eq:f2q2},\ref{eq:A2qq}) are chosen as the
largest scale of the virtuality of the internal particles in the hard $b$-quark decay diagram,
\beq
t_1=\max\{m_{B_c}\sqrt{x_2r\eta}, 1/b_1, 1/b_2\},\quad
t_2=\max\{m_{B_c}\sqrt{x_1r\eta^+}, 1/b_1, 1/b_2\}.
\eeq

%========================= reference=========================%

%%\end{CJK*}
\end{document}